\documentstyle[12pt,epsfig]{article}

\newcommand{\be}{\begin{equation}}
\newcommand{\ee}{\end{equation}}

\def\aprge{\buildrel > \over {_{\sim}}}
\begin{document}  
\topmargin 0pt
\oddsidemargin=-0.4truecm
\evensidemargin=-0.4truecm
\renewcommand{\thefootnote}{\fnsymbol{footnote}}
\newpage
\setcounter{page}{0}
\begin{titlepage}   
\vspace*{-2.0cm}  
\begin{flushright}
FISIST/01-2002/CFIF \\
hep-ph/0201089
\end{flushright}
\vspace*{0.1cm}
\begin{center}
{\Large \bf
Distinguishing magnetic moment from oscillation solutions of the 
solar neutrino problem \\
\vspace{0.16cm}
with Borexino} \\
\vspace{0.6cm}

\vspace{0.6cm}

{\large 
E. Kh. Akhmedov\footnote{On leave from National Research Centre Kurchatov 
Institute, Moscow 123182, Russia. E-mail: akhmedov@cfif.ist.utl.pt} and
Jo\~{a}o Pulido\footnote{E-mail: pulido@cfif.ist.utl.pt}}\\  
\vspace{0.15cm}
{\em Centro de F\'\i sica das Interac\c c\~oes Fundamentais (CFIF)} \\
{\em Departamento de F\'\i sica, Instituto Superior T\'ecnico }\\
{\em Av. Rovisco Pais, P-1049-001 Lisboa, Portugal}\\  
\end{center}
\vglue 0.6truecm
\begin{abstract}
Assuming that the observed deficit of solar neutrinos is due to the 
interaction of their transition magnetic moment with the solar magnetic 
field we derive the predictions for the forthcoming Borexino experiment. 
Three different model magnetic field profiles which give very good  
global fits of the currently available solar neutrino data are used.      
The expected signal at Borexino is significantly lower than those predicted 
by the LMA, LOW and VO neutrino oscillation solutions of the solar
neutrino problem. It is similar to that of the SMA oscillation solution
which, however, is strongly disfavoured by the Super-Kamiokande data on day 
and night spectra and zenith angle distribution of the events. Thus, the 
neutrino magnetic moment solution of the solar neutrino problem can be 
unambiguously distinguished from the currently favoured oscillation
solutions at Borexino.
\end{abstract}
\vspace{.5cm}
\end{titlepage}   
\renewcommand{\thefootnote}{\arabic{footnote}}
\setcounter{footnote}{0}
\section{Introduction}

If lepton flavour is not conserved, neutrinos must have flavour-off-diagonal 
(transition) magnetic moments, which applies to both Dirac and Majorana
neutrinos. Under a transverse magnetic field, such 
magnetic moments will cause a simultaneous rotation of neutrino spin and 
flavour, spin-flavour precession \cite{SV,VVO}. This precession can be 
resonantly enhanced in matter \cite{LM,Akh1,rev}, very much similarly to the 
resonance amplification of neutrino oscillations, the MSW effect \cite{MSW}.   

The resonance spin-flavour precession (RSFP) of solar neutrinos due to the 
interaction of their transition magnetic moments with the solar magnetic 
field can account for the observed deficit of solar neutrinos. The conversion 
mechanism is neutrino energy dependent, which is a necessary feature to fit 
the data.  RSFP requires relatively large values of the neutrino transition 
magnetic moment, $\mu_\nu \sim 10^{-11}\mu_B$ for peak values of the solar 
magnetic field $B_0\sim 100$ kG. Although such values of $\mu_\nu$ are not 
experimentally excluded, they are hard to achieve in the simplest extensions 
of the standard electroweak model. Still, the RSFP mechanism yields an 
excellent fit of all currently available solar neutrino data (see, e.g., 
\mbox{[7 -- 13]} for recent analyses), typically even somewhat better than 
does the large mixing angle (LMA) oscillation solution, which is the best
one among the oscillation solutions. In any case, in pursuit of the solution 
of the solar neutrino problem it is very important to test all non-standard
hypotheses, and neutrino magnetic moment seems to be the most plausible 
alternative to neutrino oscillations.  

As non-vanishing neutrino transition magnetic moments 
imply lepton flavour violation, they must be accompanied by the
usual lepton flavour mixing. Thus RSFP should in general coexist with 
neutrino oscillations. It is quite possible, however, that the flavour 
mixing in the solar neutrino sector is too small to be of any relevance to 
the solar neutrino problem. This is our assumption in the present paper, 
i.e. we neglect neutrino oscillations and consider pure RSFP transitions. 
Small flavor mixing in the solar sector does not contradict the large mixing 
in the atmospheric neutrino sector -- the corresponding mixing  angles are
independent parameters. In this connection, one can recall that the lepton 
mixing angle $\theta_{13}$ probed in short-baseline reactor neutrino
experiments is known to be small or vanishing \cite{chooz} even though the 
``atmospheric'' mixing angle $\theta_{23}$ is large \cite{atm}. 

Unfortunately, the RSFP solution of the solar neutrino problem is
difficult to establish experimentally. 
Except for predicting reduced detection rates of solar neutrinos (which the 
oscillation solutions also predict), it has mostly negative signatures:
No time variations beyond the usual $1/R^2$ variation due to the eccentricity 
of the Earth's orbit (assuming that the strongest component of the solar
magnetic field does not vary with time)\footnote{There is a caveat here which 
we shall discuss in Section 3 -- strictly speaking, this is only true when
the solar magnetic field is spherically symmetric.}; no day-night effect; no 
significant distortions of the solar neutrino spectrum in Super-Kamiokande 
and SNO experiments. One might therefore think that the RSFP solution of the 
solar neutrino problem can only be established if all the oscillation 
solutions are experimentally ruled out. Such a ``negative'' confirmation 
would hardly satisfy anyone.

In the present paper we show that in fact this is not the case: 
the RSFP predictions for the Borexino experiment are very different from 
those of neutrino oscillations, and different solutions of the solar neutrino 
problem can therefore be unambiguously distinguished experimentally. 

\section{Predictions for Borexino}
The Borexino experiment at Gran Sasso \cite{Borexino}, due to start data 
taking this year, will detect solar neutrinos through the elastic $\nu e$ 
scattering. Extremely high radiopurity of the liquid scintillator used and 
very low background will allow the detection of record low energy recoil 
electrons. In the electron kinetic energy window $T_e=250$ -- 800 keV which 
will be used in the  experiment, the major contribution to the signal ($78\%$) 
is expected from a monochromatic line of $^7$Be neutrinos with the energy 
863 keV. The next important contributions are from $^{15}$O, $^{13}$N and 
$pep$ neutrinos ($10\%$, $7.2\%$ and $3.6\%$ respectively), and the predicted 
detection rate is 55 events/day \cite{Borexino}, all according to the BP00 
standard solar model \cite{BP2000,Bahc} and assuming that neutrinos are 
``standard'' (i.e. have no mass, mixing and/or magnetic moment). A lower
signal is expected if neutrinos undergo RSFP or oscillations. 

We have calculated the expected event rates at Borexino in the case of the 
RSFP mechanism assuming that neutrinos have Majorana-like transition 
magnetic moments $\mu_\nu$ which cause the transitions $\nu_{eL}\to 
\bar{\nu}_{\mu R}$ or $\nu_{eL}\to \bar{\nu}_{\tau R}$ in the solar magnetic 
field. We have restricted ourselves to the Majorana neutrino case because it 
gives much better a fit of the solar neutrino data than the Dirac case does. 
The transition probability depends crucially on the shape and strength 
of the solar magnetic field which are essentially unknown; one therefore is
forced to use various model magnetic field profiles. In our previous work 
\cite{PA,AP,P} we have studied eight different magnetic field profiles. All 
of them except three gave either very poor or marginal global fits of the 
data of the Homestake, Gallex/GNO, SAGE, Super-Kamiokande and SNO solar
neutrino experiments \cite{data}, while the above mentioned three profiles
gave very good global fits of the data (see Table 1 below). In the present 
paper we use these three profiles to predict the signal at Borexino. We
believe that they provide a representative sample of the profiles that are 
capable of fitting the solar neutrino data. The calculation 
and fitting procedures are described in detail in our previous papers 
\cite{PA}, \cite{AP} and \cite{P}. Profiles I and II used here are profiles 
1 and 6 of ref. \cite{PA}, whereas profile III of the present paper is 
profile 4 of ref. \cite{P}. The value of $\mu_\nu$ was fixed at $10^{-11}
\mu_B$; since only the product of the magnetic moment and magnetic field 
enters in the neutrino evolution equation, our results apply to any other 
value of ${\mu_{\nu}}$ provided that the magnetic field is rescaled 
accordingly. We use the ``BP00 + new $^8$B'' standard solar model, i.e. 
the solar matter distribution and all the neutrino fluxes except the $^8$B  
one from \cite{BP2000}, whereas for the $^8$B flux we use the new value 
\cite{Bahc,BGGPG} which is based on a recent precise measurement of the 
cross section of the reaction $^7$Be$(p,\gamma)$$^8$B \cite{Jung}. 
It is about 17\% higher than the previously used value. 
We have also calculated the Borexino event rates with the ``old'' $^8$B
flux and found no significant changes in the results. 

\begin{table}[h]
{\small Table I: Reduced rates (event rates assuming RSFP divided by the
standard solar model predictions with no flavour changes allowed) for 
Gallex/GNO + SAGE, Homestake, Super-Kamiokande and SNO experiments which 
correspond to the global best fits for the three magnetic field profiles 
used, along with the corresponding  experimental data.  The values of
$\chi_{min}^2$ correspond to 39 d.o.f. See the text for more details.}
\begin{center}
\begin{tabular}{cccccccc} \hline \hline
Profile & $R_{\rm Ga}$ & $R_{\rm Cl}$ &  $R_{\rm SK}$  & $R_{\rm SNO}$ & 
$\Delta m^2$ (eV$^2$) & $B_0$ (kG) & $\chi_{min}^2$ \\
\hline
I   & 0.59 & 0.30 & 0.41 & 0.35 & $7.65\cdot 10^{-9}$ & 45 & 37.8 \\
II  & 0.58 & 0.30 & 0.39 & 0.33 & $1.60\cdot 10^{-8}$  & 113 & 36.1 \\
III & 0.58 & 0.30 & 0.40 & 0.33 & $1.48\cdot 10^{-8}$ & 101 & 35.5 \\
 Exp. & $0.57\pm .039$ & $0.30 \pm .026$ & $0.39 \pm .014$ & $0.30\pm
.025$ \\ 
\hline
\end{tabular}
\end{center}
\end{table}


In Table 1 we give, for the three magnetic field profiles used, the 
calculated reduced detection rates $R_i$ (rates assuming RSFP divided by 
those for ``standard'' neutrinos) for SAGE+Gallex/GNO (Ga), Homestake (Cl), 
Super-Kamiokande (SK) and SNO experiments. The indicated rates correspond 
to the best global fits of the data (all rates plus day and night spectra 
at Super-Kamiokande). For each profile we show the corresponding best-fit  
values of $\Delta m^2$, peak magnetic field strength parameter $B_0$, and 
$\chi_{min}^2$ (39 d.o.f.). In the last line we give the experimental 
detection rates normalized to the ``BP00 + new $^8$B'' standard solar model.   

As can be seen from the table, all three profiles yield very good global
fits of the data. Profile I produces slightly worse a fit than those given 
by profiles II and III, mainly because it predicts too high a SNO rate. 
For the ``old'' values of the $^8$B neutrino flux, the allowed regions of 
parameters at 95\% CL and 99\% CL for profiles I and II and for profile III 
 were given in fig. 1 and fig. 2 of ref. \cite{P}, respectively. If one 
uses instead the new $^8$B flux, the allowed regions are slightly shifted 
(by about 5\%) towards higher values of the magnetic field strengths. 
For illustration, in fig. 1 we show the 95\% CL allowed regions for profile
III for both old and new $^8$B fluxes. 
   
\begin{figure}[h]
\setlength{\unitlength}{1cm}
\begin{center}
\hspace*{-1.6cm}
\epsfig{file=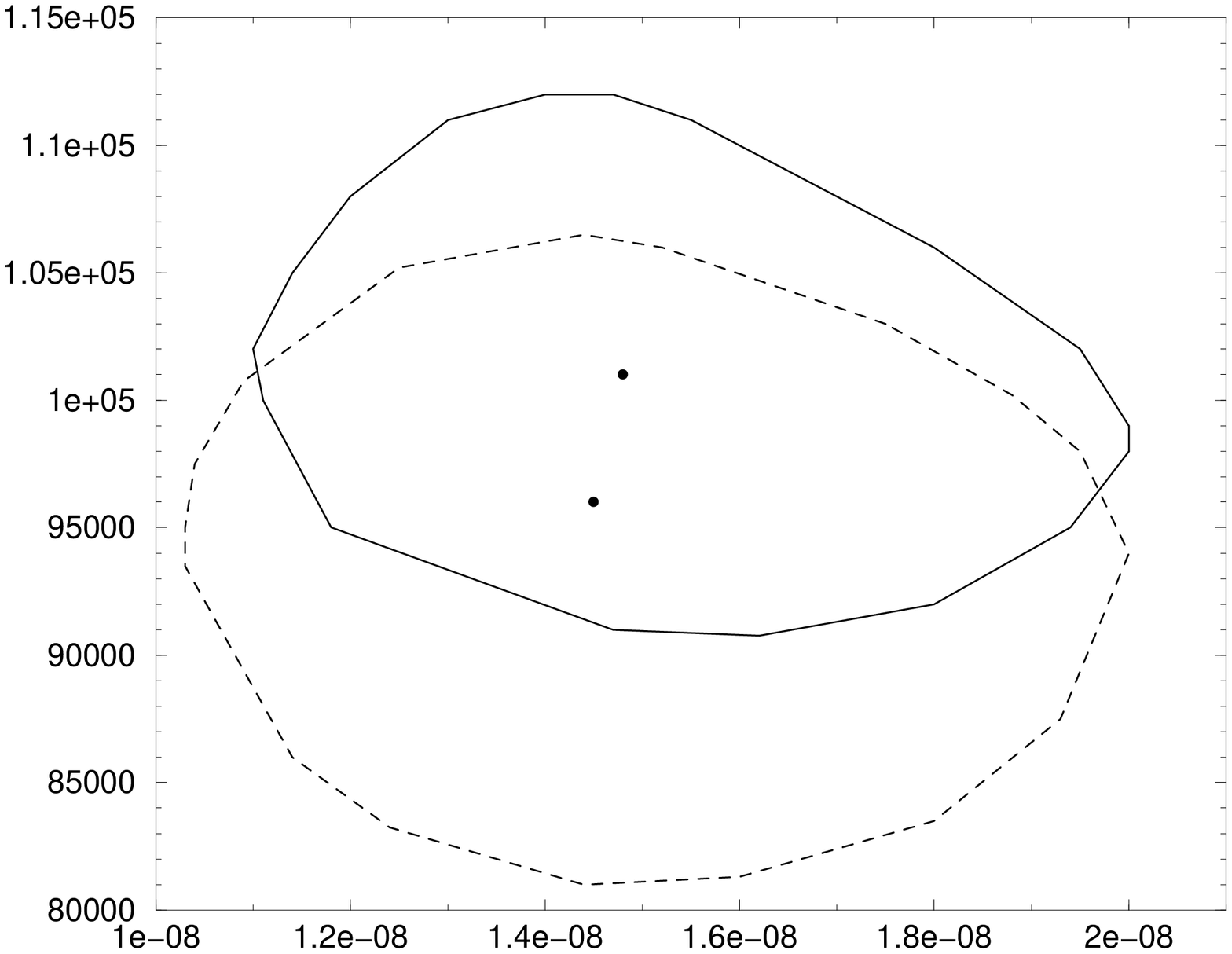,height=8.0cm,angle=270}
\end{center}
\vspace{-0.2cm}
{\small Figure 1: Allowed regions of $\Delta m^2$ (in eV$^2$) and $B_0$
(in G) for profile III at 95\% CL. Solid curve -- new $^8$B flux, dashed 
curve -- old $^8$B flux. The best fit points are shown by dots.}
\end{figure}

In Table 2 predictions are given for the reduction factors of the individual 
contributions of various solar neutrino fluxes to the Borexino event rate  
(the contributions to the event rate assuming RSFP divided by those for 
``standard'' neutrinos). The results are given, for each of the three magnetic 
field profiles, for the values of $\Delta m^2$ and $B_0$ that produced the 
best global fits of the currently available data (see Table 1). 

\begin{table}
{\small Table II: Columns 2 to 8 -- reduction factors of the individual 
contributions of different solar neutrino fluxes to the Borexino event rate 
(the contributions to the event rate assuming RSFP divided by those for 
``standard'' neutrinos). Last column -- reduction factors for the total 
event rate. The values of $\Delta m^2$ and $B_0$ correspond to the best 
fits of the present data (see Table 1).}
\begin{center}
\begin{tabular}{ccccccccc} \hline \hline
Profile & $pp$ & $pep$ & $^7$Be  & $^{15}$O & $^{13}$N & $^8$B &
$hep$ & total \\
\hline
I   & 0.71 & 0.18 & 0.29 & 0.26 & 0.32 & 0.44 & 0.47 & 0.28 \\
II  & 0.64 & 0.24 & 0.42 & 0.35 & 0.43 & 0.42 & 0.44 & 0.41 \\
III & 0.62 & 0.30 & 0.35 & 0.35 & 0.39 & 0.44 & 0.46 & 0.35 \\  
\hline 
\end{tabular}
\end{center}
\end{table}

Finally, in Table III we present the predicted values of the reduced event 
rates for Borexino $R_{\rm Bor}$. We give there the values of $R_{\rm Bor}$ 
corresponding to the best-fit values of $\Delta m^2$ and $B_0$ 
as well as the minimum and maximum values of $R_{\rm Bor}$ corresponding to 
the 95\% CL and 99\% CL allowed ranges of $\Delta m^2$ and $B_0$. 


\begin{table}
{\small Table III: Predicted reduced event rates (rates assuming RSFP divided 
by the standard solar model predictions with no flavour changes allowed) for 
Borexino $R_{\rm Bor}$.}
\begin{center}
\begin{tabular}{cccccc} \hline \hline
Profile & b.f.& min (95\% CL)&  max (95\% CL)& min (99\% CL)& max (99\% CL)\\
\hline
I   & 0.28 & 0.21 & 0.50 & 0.21 & 0.57 \\
II  & 0.41 & 0.29 & 0.57 & 0.28 & 0.62 \\
III & 0.35 & 0.31 & 0.52 & 0.30 & 0.57 \\  
\hline
\end{tabular}
\end{center}
\end{table}


\section{Discussion}
The main goal of this work was to investigate whether RSFP can be 
distinguished from the oscillation solutions of the solar neutrino problem  
at Borexino. There are four main types of oscillation solutions of the 
solar neutrino problem, depending on the allowed values of the leptonic mixing 
angle $\theta$ and neutrino mass squared difference $\Delta m^2$: Large mixing 
angle (LMA), small mixing angle (SMA) and low--$\Delta m^2$ (LOW) MSW 
solutions, and also vacuum oscillation (VO) solution (for recent discussions 
see, e.g., \cite{Lisi2001,KS2001,BGGPG,GG,aliani,Gago}). 
The LMA, LOW and VO solutions all predict the average suppression of the 
event rate at Borexino by 35 -- 40\%, whereas in the case of the SMA
solution a suppression by about a factor of five is expected. 
 
The main feature of the RSFP mechanism which can be exploited in order to 
distinguish it experimentally from neutrino oscillations is the peculiar 
shape of the energy dependence of the survival probability of solar neutrinos: 
At high energies it resembles the $\nu_e$ survival probability of the LMA 
oscillation solution, whereas at low energies it is similar to that of the 
SMA solution. A mismatch in the results of the experiments sensitive to
the high-energy and low-energy parts of the solar neutrino spectrum would 
therefore be an indication for RSFP.

As can be seen from Table III, the RSFP mechanism predicts the suppression 
of the event rate at Borexino by about a factor of three. The maximum allowed 
at 99\% CL reduced rate is 0.62; this only marginally overlaps with the
minimum allowed at 3$\sigma$ reduced rate in the case of the LMA solution 
(0.58, see Table 7 of ref. \cite{BGGPG}). Thus, the predictions of 
the RSFP and LMA solutions are more than 5$\sigma$ away from each other and 
the probability of mistaking one for another is very low. 

The minimum allowed at 3$\sigma$ values of the reduced rate at Borexino in
the case of LOW and VO solutions, 0.54 and 0.53 respectively \cite{BGGPG}, 
are slightly lower than that for the LMA solution, so that there is a
larger overlap with the 99\% CL prediction of the RSFP. However, in these 
cases, too, one can easily discriminate between RSFP and the oscillation 
solutions. Indeed, in the case of the LOW  solution one expects a sizeable
(up to 40\%) day-night event rate difference at Borexino, while VO should
lead to large seasonal variations beyond the usual $1/R^2$ dependence. No 
such effects are predicted by RSFP. 

Our predictions for the reduced event rate at Borexino in the case of 
RSFP are slightly higher than those of the SMA oscillation solution, although 
there is a significant overlap between the predicted rates in these two cases. 
It should be noted, however, that the SMA solution is strongly disfavoured by 
the data on day and night spectra and zenith angle distributions of recoil 
electrons at Super-Kamiokande \cite{SK}. We therefore conclude that Borexino 
will allow a clear discrimination between RSFP and currently favoured 
oscillation solutions of the solar neutrino problem. It should be noted that 
new dedicated low-energy solar neutrino experiments, which are widely 
discussed now \cite{lownu}, should have a similar or even stronger
discriminative power 
\footnote{We thank M. Nakahata for pointing this out to us.}.  

The RSFP mechanism may also lead to some specific effects, absent in the
case of neutrino oscillation solutions. If the solar magnetic field
is not axially symmetric, the rotation of the Sun can lead to a time 
variation of the signal with the period equal to the solar rotation
period (about 28 Earth's days). 
Seasonal variations of the signal can also occur due to the inclination 
(by about $7^\circ$) of the solar equatorial plane to the Earth's orbit, 
provided that the solar magnetic field depends on the polar angle $\Theta$. 
This effect depends on the three-dimensional structure of the solar magnetic 
field. For the model profile of ref. \cite{Mir1}, the transverse component
$B_\perp \propto \sin\Theta$; since for solar neutrinos reaching the Earth
$\Theta=90^\circ \pm 7^\circ$, one finds seasonal variations of less 
than $\pm 1.5\%$ for charged-current signals. In the case of neutrino 
detection through $\nu e$ scattering (Super-Kamiokande, SNO and Borexino), 
these variations are further diluted by the neutral-current contribution
to the event rates. Thus, the seasonal variations of this kind are probably 
too small to be observable. 

Another possible signature of RSFP is an 
observable flux of $\bar{\nu}_e$ from the Sun if neutrinos, in addition to
transition magnetic moments, have a sizeable flavour mixing ($\theta \aprge 
0.1$) \cite{antinu}. The flux of solar $\bar{\nu}_e$'s at the level of 1\%
of the $\nu_e$ flux can, in principle, be detected at 
Borexino and SNO. However, these signatures depend on additional
assumptions about $\theta$ and the structure of the solar magnetic field,   
whereas our predictions for the Borexino detection rate are essentially 
model independent. The only possible model dependence is contained in the 
choice of the solar magnetic field profile, and this freedom is
severely constrained by the requirement of fitting the available solar
neutrino  data. As a result, the predictions for the Borexino event rate, 
though somewhat different for different profiles (see Table III), all
fall below those for the LMA, LOW and VO solutions.

In conclusion, we have shown that the Borexino experiment will be able to 
unambiguously distinguish RSFP from the currently favoured oscillation 
solutions of the solar neutrino problem. 

\vspace{0.1cm}  
\noindent

{\em Acknowledgements.} We are grateful to M. Nakahata for useful
correspondence. E.A. was supported by the Calouste Gulbenkian
Foundation as a Gulbenkian Visiting Professor at Instituto Superior 
T\'ecnico.

\end{document}